\pdfoutput=1

\documentclass{webofc}
\usepackage[varg]{txfonts}   

\usepackage{overpic}
\usepackage{relsize}
\usepackage{alumisc}
\usepackage{ifthen}
\usepackage{hyperref}
\usepackage{booktabs}
\usepackage[svgnames,x11names]{xcolor}
\usepackage{tikz}
\usepackage{amsmath}
\usepackage{amssymb}
\usepackage{amstext}
\usepackage{varwidth}

{\(\displaystyle}
{\)}

\newif\ifhevea\heveafalse

\input{htdef.tex}
\input{babarsym}
\def\taumtomu   {\ensuremath{\taum \to \mun \numb \nut}\xspace}

\def\taumtotnpiz   {\ensuremath{\taum \to t^{-} n\piz \nut}\xspace}

\def\taumtopinpiz   {\ensuremath{\taum \to \pim n \piz \nut}\xspace}

\def\dpinpiz   {\ensuremath{\pim n \piz \nut}\xspace}

\def\taumtoKnpiz   {\ensuremath{\taum \to \Km n \piz \nut}\xspace}

\def\dKnpiz        {\ensuremath{\Km n \piz \nut}\xspace}

\def\taumtopi   {\ensuremath{\taum \to \pim \nut}\xspace}

\def\taumtopipiz   {\ensuremath{\taum \to \pim \piz \nut}\xspace}

\def\taumtopitwopiz   {\ensuremath{\taum \to \pim 2\piz \nut}\xspace}

\def\taumtopithreepiz   {\ensuremath{\taum \to \pim 3\piz \nut}\xspace}

\def\dpithreepiz   {\ensuremath{\pim 3\piz}\xspace}

\def\taumtopifourpiz   {\ensuremath{\taum \to \pim 4\piz \nut}\xspace}

\def\dpifourpiz                  {\ensuremath{\pim 4\piz }\xspace}

\def\taumtopifivepiz   {\ensuremath{\taum \to \pim 5\piz \nut}\xspace}

\def\taumtoK   {\ensuremath{\taum \to \Km \nut}\xspace}

\def\dK        {\ensuremath{\Km}\xspace}

\def\taumtoKpiz   {\ensuremath{\taum \to \Km \piz \nut}\xspace}

\def\dKpiz        {\ensuremath{\Km \piz }\xspace}

\def\taumtoKtwopiz   {\ensuremath{\taum \to \Km 2\piz \nut}\xspace}

\def\dKtwopiz                  {\ensuremath{\Km 2\piz }\xspace}

\def\taumtoKthreepiz   {\ensuremath{\taum \to \Km 3\piz \nut}\xspace}

\def\dKthreepiz        {\ensuremath{\Km 3\piz}\xspace}

\def\taumtoKfourpiz   {\ensuremath{\taum \to \Km 4\piz \nut}\xspace}

\def\taumtoKKzpiz   {\ensuremath{\taum \to \Km \Kzb \piz \nut}\xspace}

\def\taumtopipipi   {\ensuremath{\taum \to \pim \pip \pim \nut}\xspace}

\def\taumtoKKpi   {\ensuremath{\taum \to \pim \Kp \Km  \nut}\xspace}

\def\pc {\ensuremath{\%}\xspace }

\def\geomCenter{\mbox{C\kern-0.4em\raisebox{-2pt}{\smaller L} }}

\def\L     {\ensuremath{\mathcal{L}}\xspace}

\def\photos      {\mbox{\tt PHOTOS}\xspace}
\def\kkmc        {\mbox{\tt KKMC}\xspace}
\def\jetset      {\mbox{\tt Jetset}\xspace}
\def\geantFour   {\mbox{\tt GEANT4}\xspace}
\def\tauola      {\mbox{\tt TAUOLA}\xspace}

\newcommand{\tabRef}[1]    {Table~\ref{#1}}

\newcommand{\EE}[1]{\ensuremath{\times 10^{#1}}}

\makeatletter
\providecommand{\htdef}[2]{%
  \@namedef{hfagtau@#1}{#2}%
}

\providecommand{\htuse}[1]{%
  \ifcsname hfagtau@#1\endcsname
  \@nameuse{hfagtau@#1}%
  \else
  \@latex@error{Undefined name hfagtau@#1}\@eha
  \fi
}
\makeatother

\htdef{BrK_val}{7.174}%
\htdef{BrK_stat}{0.033}%
\htdef{BrK_syst}{0.213}%
\htdef{BrK_ord}{-3}%
\htdef{BrK_statsyst}{(\htuse{BrK_val} \pm \htuse{BrK_stat} \pm \htuse{BrK_syst})\EE{\htuse{BrK_ord}}}
\htdef{BrKPi0_val}{5.054}%
\htdef{BrKPi0_stat}{0.021}%
\htdef{BrKPi0_syst}{0.148}%
\htdef{BrKPi0_ord}{-3}%
\htdef{BrKPi0_statsyst}{(\htuse{BrKPi0_val} \pm \htuse{BrKPi0_stat} \pm \htuse{BrKPi0_syst})\EE{\htuse{BrKPi0_ord}}}
\htdef{BrK2Pi0_val}{6.151}%
\htdef{BrK2Pi0_stat}{0.117}%
\htdef{BrK2Pi0_syst}{0.338}%
\htdef{BrK2Pi0_ord}{-4}%
\htdef{BrK2Pi0_statsyst}{(\htuse{BrK2Pi0_val} \pm \htuse{BrK2Pi0_stat} \pm \htuse{BrK2Pi0_syst})\EE{\htuse{BrK2Pi0_ord}}}
\htdef{BrK3Pi0_val}{1.246}%
\htdef{BrK3Pi0_stat}{0.164}%
\htdef{BrK3Pi0_syst}{0.238}%
\htdef{BrK3Pi0_ord}{-4}%
\htdef{BrK3Pi0_statsyst}{(\htuse{BrK3Pi0_val} \pm \htuse{BrK3Pi0_stat} \pm \htuse{BrK3Pi0_syst})\EE{\htuse{BrK3Pi0_ord}}}
\htdef{BrPi3Pi0_val}{1.168}%
\htdef{BrPi3Pi0_stat}{0.006}%
\htdef{BrPi3Pi0_syst}{0.038}%
\htdef{BrPi3Pi0_ord}{-2}%
\htdef{BrPi3Pi0_statsyst}{(\htuse{BrPi3Pi0_val} \pm \htuse{BrPi3Pi0_stat} \pm \htuse{BrPi3Pi0_syst})\EE{\htuse{BrPi3Pi0_ord}}}
\htdef{BrPi4Pi0_val}{9.020}%
\htdef{BrPi4Pi0_stat}{0.400}%
\htdef{BrPi4Pi0_syst}{0.652}%
\htdef{BrPi4Pi0_ord}{-4}%
\htdef{BrPi4Pi0_statsyst}{(\htuse{BrPi4Pi0_val} \pm \htuse{BrPi4Pi0_stat} \pm \htuse{BrPi4Pi0_syst})\EE{\htuse{BrPi4Pi0_ord}}}
\htdef{central_val_abs}{7.174 & 5.054 & 6.151 & 1.246 & 1.168 & 9.020}%
\htdef{tot_uncert_abs}{0.216 & 0.149 & 0.357 & 0.289 & 0.038 & 0.765}%
\htdef{stat_uncert_abs}{0.033 & 0.021 & 0.117 & 0.164 & 0.006 & 0.400}%
\htdef{syst_uncert_abs}{0.213 & 0.148 & 0.338 & 0.238 & 0.038 & 0.652}%
\htdef{central_val}{100.00 & 100.00 & 100.00 & 100.00 & 100.00 & 100.00}%
\htdef{tot_uncert}{3.00 & 2.95 & 5.81 & 23.20 & 3.27 & 8.48}%
\htdef{stat_uncert}{0.46 & 0.41 & 1.91 & 13.13 & 0.52 & 4.44}%
\htdef{syst_uncert}{2.97 & 2.93 & 5.49 & 19.13 & 3.23 & 7.23}%
\htdef{signalEffi}{0.27 & 0.27 & 0.87 & 3.99 & 0.27 & 1.50}%
\htdef{bkgEffi}{0.15 & 0.15 & 0.87 & 6.32 & 0.11 & 1.67}%
\htdef{bkgNorm}{0.18 & 0.30 & 1.44 & 11.52 & 0.21 & 3.49}%
\htdef{pidTable}{0.15 & 0.11 & 0.18 & 0.71 & 0.08 & 0.20}%
\htdef{pidCorr}{1.83 & 1.55 & 1.78 & 2.56 & 0.20 & 0.26}%
\htdef{lumi}{0.79 & 0.93 & 1.40 & 2.62 & 0.71 & 0.98}%
\htdef{trackEffi}{0.43 & 0.50 & 0.76 & 1.42 & 0.38 & 0.53}%
\htdef{splitoff}{1.52 & 1.84 & 2.77 & 5.18 & 1.40 & 1.94}%
\htdef{pi0Cor}{0.03 & 1.20 & 3.63 & 10.56 & 2.76 & 5.36}%
\htdef{downFeedP}{0.00 & 0.00 & 0.00 & 0.02 & 0.04 & 1.08}%
\htdef{downFeedK}{0.00 & 0.00 & 0.13 & 4.78 & 0.00 & 0.00}%
\htdef{muMisID}{1.48 & 0.01 & 0.00 & 0.00 & 0.00 & 0.00}%
\htdef{stat_corr_mat}{%
 & \ensuremath{K} & \ensuremath{K\piz} & \ensuremath{K2\piz} & \ensuremath{K3\piz} & \ensuremath{\pi3\piz} & \ensuremath{\pi4\piz} \\
\ensuremath{K} & 1.000 & -0.029 & 0.001 & -0.000 & -0.000 & 0.000 \\
\ensuremath{K\piz} & -0.029 & 1.000 & -0.086 & 0.004 & -0.000 & -0.000 \\
\ensuremath{K2\piz} & 0.001 & -0.086 & 1.000 & -0.208 & -0.002 & 0.002 \\
\ensuremath{K3\piz} & -0.000 & 0.004 & -0.208 & 1.000 & -0.038 & -0.005 \\
\ensuremath{\pi3\piz} & -0.000 & -0.000 & -0.002 & -0.038 & 1.000 & -0.312 \\
\ensuremath{\pi4\piz} & 0.000 & -0.000 & 0.002 & -0.005 & -0.312 & 1.000 \\}%
\htdef{syst_corr_mat}{%
 & \ensuremath{K} & \ensuremath{K\piz} & \ensuremath{K2\piz} & \ensuremath{K3\piz} & \ensuremath{\pi3\piz} & \ensuremath{\pi4\piz} \\
\ensuremath{K} & 1.000 & 0.743 & 0.506 & 0.251 & 0.299 & 0.190 \\
\ensuremath{K\piz} & 0.743 & 1.000 & 0.859 & 0.554 & 0.720 & 0.542 \\
\ensuremath{K2\piz} & 0.506 & 0.859 & 1.000 & 0.624 & 0.875 & 0.684 \\
\ensuremath{K3\piz} & 0.251 & 0.554 & 0.624 & 1.000 & 0.636 & 0.529 \\
\ensuremath{\pi3\piz} & 0.299 & 0.720 & 0.875 & 0.636 & 1.000 & 0.805 \\
\ensuremath{\pi4\piz} & 0.190 & 0.542 & 0.684 & 0.529 & 0.805 & 1.000 \\}%
\htdef{tot_corr_mat}{%
 & \ensuremath{K} & \ensuremath{K\piz} & \ensuremath{K2\piz} & \ensuremath{K3\piz} & \ensuremath{\pi3\piz} & \ensuremath{\pi4\piz} \\
\ensuremath{K} & 1.000 & 0.726 & 0.472 & 0.205 & 0.292 & 0.160 \\
\ensuremath{K\piz} & 0.726 & 1.000 & 0.799 & 0.452 & 0.704 & 0.458 \\
\ensuremath{K2\piz} & 0.472 & 0.799 & 1.000 & 0.448 & 0.816 & 0.551 \\
\ensuremath{K3\piz} & 0.205 & 0.452 & 0.448 & 1.000 & 0.514 & 0.370 \\
\ensuremath{\pi3\piz} & 0.292 & 0.704 & 0.816 & 0.514 & 1.000 & 0.651 \\
\ensuremath{\pi4\piz} & 0.160 & 0.458 & 0.551 & 0.370 & 0.651 & 1.000 \\}%

\htdef{br_TauToPi_wa_val}{10.828}%
\htdef{br_TauToPi_wa_unc}{0.105}%
\htdef{br_TauToPi_wa_val_unc_fmt}{\ensuremath{10.828\hphantom{0} \pm \hphantom{0}0.105\hphantom{0}}}%
\htdef{br_TauToPiPi0_wa_val}{25.46}%
\htdef{br_TauToPiPi0_wa_unc}{0.12}%
\htdef{br_TauToPiPi0_wa_val_unc_fmt}{\ensuremath{25.46\hphantom{00} \pm \hphantom{0}0.12\hphantom{00}}}%
\htdef{br_TauToPi2Pi0_wa_val}{9.239}%
\htdef{br_TauToPi2Pi0_wa_unc}{0.124}%
\htdef{br_TauToPi2Pi0_wa_val_unc_fmt}{\ensuremath{9.239\hphantom{0} \pm \hphantom{0}0.124\hphantom{0}}}%
\htdef{br_TauToPi3Pi0_wa_val}{0.977}%
\htdef{br_TauToPi3Pi0_wa_unc}{0.09}%
\htdef{br_TauToPi3Pi0_wa_val_unc_fmt}{\ensuremath{0.977\hphantom{0} \pm \hphantom{0}0.09\hphantom{00}}}%
\htdef{br_TauToPi4Pi0_wa_val}{0.112}%
\htdef{br_TauToPi4Pi0_wa_unc}{0.051}%
\htdef{br_TauToPi4Pi0_wa_val_unc_fmt}{\ensuremath{0.112\hphantom{0} \pm \hphantom{0}0.051\hphantom{0}}}%
\htdef{br_TauToPi5Pi0_wa_val}{NA}%
\htdef{br_TauToPi5Pi0_wa_unc}{NA}%
\htdef{br_TauToPi5Pi0_wa_val_unc_fmt}{\ensuremath{NA\hphantom{00} \pm NA\hphantom{00}}}%
\htdef{br_TauToK_wa_val}{0.685}%
\htdef{br_TauToK_wa_unc}{0.023}%
\htdef{br_TauToK_wa_val_unc_fmt}{\ensuremath{0.685\hphantom{0} \pm \hphantom{0}0.023\hphantom{0}}}%
\htdef{br_TauToKPi0_wa_val}{0.426}%
\htdef{br_TauToKPi0_wa_unc}{0.016}%
\htdef{br_TauToKPi0_wa_val_unc_fmt}{\ensuremath{0.426\hphantom{0} \pm \hphantom{0}0.016\hphantom{0}}}%
\htdef{br_TauToK2Pi0_wa_val}{0.058}%
\htdef{br_TauToK2Pi0_wa_unc}{0.024}%
\htdef{br_TauToK2Pi0_wa_val_unc_fmt}{\ensuremath{0.058\hphantom{0} \pm \hphantom{0}0.024\hphantom{0}}}%
\htdef{br_TauToK3Pi0_wa_val}{0.037}%
\htdef{br_TauToK3Pi0_wa_unc}{0.024}%
\htdef{br_TauToK3Pi0_wa_val_unc_fmt}{\ensuremath{0.037\hphantom{0} \pm \hphantom{0}0.024\hphantom{0}}}%
\htdef{br_TauToK4Pi0_wa_val}{NA}%
\htdef{br_TauToK4Pi0_wa_unc}{NA}%
\htdef{br_TauToK4Pi0_wa_val_unc_fmt}{\ensuremath{NA\hphantom{00} \pm NA\hphantom{00}}}%
\htdef{br_TauToKEta_wa_val}{0.0154}%
\htdef{br_TauToKEta_wa_unc}{0.0008}%
\htdef{br_TauToKEta_wa_val_unc_fmt}{\ensuremath{0.0154 \pm \hphantom{0}0.0008}}%
\htdef{br_TauToPiEtaPi0_wa_val}{0.138}%
\htdef{br_TauToPiEtaPi0_wa_unc}{0.009}%
\htdef{br_TauToPiEtaPi0_wa_val_unc_fmt}{\ensuremath{0.138\hphantom{0} \pm \hphantom{0}0.009\hphantom{0}}}%
\htdef{br_TauToPiEta2Pi0_wa_val}{0.0181}%
\htdef{br_TauToPiEta2Pi0_wa_unc}{0.0031}%
\htdef{br_TauToPiEta2Pi0_wa_val_unc_fmt}{\ensuremath{0.0181 \pm \hphantom{0}0.0031}}%
\htdef{br_TauToKEtaPi0_wa_val}{0.0048}%
\htdef{br_TauToKEtaPi0_wa_unc}{0.0012}%
\htdef{br_TauToKEtaPi0_wa_val_unc_fmt}{\ensuremath{0.0048 \pm \hphantom{0}0.0012}}%
\htdef{br_TauToPiK0_wa_val}{0.839}%
\htdef{br_TauToPiK0_wa_unc}{0.022}%
\htdef{br_TauToPiK0_wa_val_unc_fmt}{\ensuremath{0.839\hphantom{0} \pm \hphantom{0}0.022\hphantom{0}}}%
\htdef{br_TauToPiK0Pi0_wa_val}{0.383}%
\htdef{br_TauToPiK0Pi0_wa_unc}{0.014}%
\htdef{br_TauToPiK0Pi0_wa_val_unc_fmt}{\ensuremath{0.383\hphantom{0} \pm \hphantom{0}0.014\hphantom{0}}}%
\htdef{br_TauToKK0_wa_val}{0.149}%
\htdef{br_TauToKK0_wa_unc}{0.005}%
\htdef{br_TauToKK0_wa_val_unc_fmt}{\ensuremath{0.149\hphantom{0} \pm \hphantom{0}0.005\hphantom{0}}}%
\htdef{br_TauToKK0Pi0_wa_val}{0.149}%
\htdef{br_TauToKK0Pi0_wa_unc}{0.007}%
\htdef{br_TauToKK0Pi0_wa_val_unc_fmt}{\ensuremath{0.149\hphantom{0} \pm \hphantom{0}0.007\hphantom{0}}}%
\htdef{br_TauToPiK0Eta_wa_val}{0.0093}%
\htdef{br_TauToPiK0Eta_wa_unc}{0.0015}%
\htdef{br_TauToPiK0Eta_wa_val_unc_fmt}{\ensuremath{0.0093 \pm \hphantom{0}0.0015}}%
\htdef{br_TauToPiK0K0_wa_val}{0.153}%
\htdef{br_TauToPiK0K0_wa_unc}{0.034}%
\htdef{br_TauToPiK0K0_wa_val_unc_fmt}{\ensuremath{0.153\hphantom{0} \pm \hphantom{0}0.034\hphantom{0}}}%
\htdef{br_TauToPimPipPim_wa_val}{9.041}%
\htdef{br_TauToPimPipPim_wa_unc}{0.097}%
\htdef{br_TauToPimPipPim_wa_val_unc_fmt}{\ensuremath{9.041\hphantom{0} \pm \hphantom{0}0.097\hphantom{0}}}%
\htdef{br_TauToKmPipPim_wa_val}{0.29}%
\htdef{br_TauToKmPipPim_wa_unc}{0.018}%
\htdef{br_TauToKmPipPim_wa_val_unc_fmt}{\ensuremath{0.29\hphantom{00} \pm \hphantom{0}0.018\hphantom{0}}}%
\htdef{br_TauToKmKpPim_wa_val}{0.143}%
\htdef{br_TauToKmKpPim_wa_unc}{0.007}%
\htdef{br_TauToKmKpPim_wa_val_unc_fmt}{\ensuremath{0.143\hphantom{0} \pm \hphantom{0}0.007\hphantom{0}}}%
\htdef{br_TauToelectron_wa_val}{17.82}%
\htdef{br_TauToelectron_wa_unc}{0.05}%
\htdef{br_TauToelectron_wa_val_unc_fmt}{\ensuremath{17.82\hphantom{00} \pm \hphantom{0}0.05\hphantom{00}}}%
\htdef{br_TauTomuon_wa_val}{17.33}%
\htdef{br_TauTomuon_wa_unc}{0.05}%
\htdef{br_TauTomuon_wa_val_unc_fmt}{\ensuremath{17.33\hphantom{00} \pm \hphantom{0}0.05\hphantom{00}}}%

\htdef{sample_lumi}{473.9}

\begin{document}
\title{Measurement of the branching fractions of the
  decays $\taumtoKnpiz$ ($n = 0,1,2,3$) and $\taumtopinpiz$ ($n = 3,4$)
  by \babar}
%
%

\author{Alberto
  Lusiani\inst{1,2}\fnsep\thanks{\email{alberto.lusiani@pi.infn.it}}}

\institute{%
  Scuola Normale Superiore, Pisa, Italy\and INFN sezione di Pisa, Pisa,
  Italy}

\abstract{%
  We report preliminary measurements of the branching fractions of the
  decays $\taumtoKnpiz$ ($n = 0,1,2,3$) and $\taumtopinpiz$ ($n = 3,4$),
  excluding the contributions that proceed through the decay of
  intermediate $K^0$ and $\eta$ mesons. The measurements are based on a
  data sample of 435 million \mtau pairs produced in \epem collisions at
  and near the $\Upsilon(4S)$ peak and collected with the \babar
  detector in 1999--2008.  The measured branching fractions are
  $\BR(\taumtoK) = \htuse{BrK_statsyst}$,
  $\BR(\taumtoKpiz) = \htuse{BrKPi0_statsyst}$,
  $\BR(\taumtoKtwopiz) = \htuse{BrK2Pi0_statsyst}$,
  $\BR(\taumtoKthreepiz) = \htuse{BrK3Pi0_statsyst}$,
  $\BR(\taumtopithreepiz) = \htuse{BrPi3Pi0_statsyst}$,
  $\BR(\taumtopifourpiz) = \htuse{BrPi4Pi0_statsyst}$, where the first
  uncertainty is statistical and the second one systematic.  }

\maketitle

\section{Introduction}

The branching fractions of the \mtau lepton into strange and non-strange
final states, respectively $\BR(\tau\to X_s\nu)$ and $\BR(\tau\to X_d\nu)$, can be
used to determine the Cabibbo-Kobayashi-Maskawa (CKM) quark mixing
matrix element \Vus~\cite{Gamiz:2002nu,Gamiz:2004ar}. The resulting \Vus
value~\cite{Amhis:2016xyh} is more than $3\sigma$ lower than the value
that is obtained from the the \Vud and \Vub measurements with the
assumption that the CKM matrix is
unitary~\cite{Patrignani:2016xqp,Amhis:2016xyh}. The experimental
uncertainty of this \Vus determination is dominated by the uncertainties
on the \mtau branching fractions into states with an odd number of
kaons, which are summed to obtain
$\BR(\tau\to X_s\nu)$~\cite{Lusiani:2018zvr}.

We report measurements of the branching fractions of the decays
\taumtoKnpiz with $n = 0,1,2,3$ and of the decays \taumtopinpiz with
$n = 3,4$. Charge conjugate decays are implied.  All measurements
exclude the decays that proceed through $K_S^0 \to 2\pi^0$ or $\eta \to 3\pi^0$
to the above final states.  These measurements
significantly improve some of the least precise experimental inputs that
are involved in the above mentioned \Vus determination.

\section{Analysis}

We analyzed \epem collisions at and near a center-of-mass (CM) energy of
$\sqrt{s} = 10.58\gev$, recorded by the \babar
detector~\cite{Aubert:2001tu} at the PEP-II asymmetric-energy storage
rings operated at the SLAC National Accelerator Laboratory. The data
sample consists in about 435 million \tautau pairs, corresponding to an
integrated luminosity $\mathcal{L} = \htuse{sample_lumi}\,\invfb$ and a
luminosity-weighted average cross-section of
$\sigma(e^+e^- \to \tau^+\tau^-) = (0.919\pm0.003)$\,nb~\cite{Jadach:1999vf,Banerjee:2007is},


The \babar\ detector is described in detail in
Refs.~\cite{Aubert:2001tu,TheBABAR:2013jta}.  Charged particles are
reconstructed as tracks with a five-layer silicon vertex detector (SVT)
and a 40-layer drift chamber (DCH) inside a $1.5\,$T magnetic field.
An electromagnetic calorimeter (EMC) comprised of 6580 CsI(Tl)
crystals is used to identify electrons and photons.  A ring-imaging
Cherenkov detector (DIRC) is used to identify charged hadrons and to
provide additional lepton identification information.  These detectors
are located inside a superconducting solenoidal magnet that produces a
1.5\,T magnetic field and whose magnetic-flux return is instrumented to
identify muons (IFR).


Monte Carlo simulated events are used to evaluate background
contamination and selection efficiencies and to study systematic
effects.  Simulated $\epem \to \tau^+\tau^-$ events are produced using
the \kkmc generator~\cite{Jadach:1999vf} and the \tauola decay
library~\cite{Jadach:1993hs}. \jetset~\cite{Sjostrand:1993yb} is used to
simulate $\epem \to \qqbar$ with $q= u, d, s, c$ and
\evtgen~\cite{Lange:2001uf} is used to simulate the decays of the $B$
mesons.  Final-state radiative effects are simulated using
\photos~\cite{Golonka:2005pn}. The detector response is simulated with
\geantFour~\cite{Agostinelli:2002hh,Allison:2006ve}. All simulated
events are reconstructed in the same manner as the data.  The number of
simulated events is comparable to the number expected in the data for
all processes, with the exception of Bhabha and two-photon events, which
are not simulated and are studied on data.


The analysis proceeds as follows. We select candidate events consisting
of \mtau pairs where one \mtau decays leptonically and the other one
decays to \dKnpiz ($n = 0,1,2,3$) and $\dpinpiz$ ($n = 3,4$), assigning
each event exclusively to a single signal mode according to the hadron
type and the number of reconstructed neutral pions.  We use the Monte
Carlo simulation to subtract the expected backgrounds and to account for
cross-feeds due to reconstruction mismatches, in order to obtain the
number of the events produced for each signal mode. Finally, we compute
the corresponding branching fractions, using the estimated number of
produced \mtau pairs.

Signal candidates are required to have two well-reconstructed
oppositely-charged tracks, whose point of closest approach to the beam
axis must be closer than 1.5\cm in the transverse plane, and closer than
2.5\cm along the beam axis to the interaction region center. To insure
good particle identification (PID), tracks must be within the EMC and
DIRC acceptance and have a transverse momentum greater than 0.25\gevc to
ensure that they reach the DIRC. Tracks are assigned to one of two
hemispheres according to the sign of their projection onto the event
thrust axis~\cite{Brandt:1964sa}, computed using tracks and EMC energy
deposits with energy $E>50\mev$.  The two tracks must belong to opposite
hemispheres.

The tracking devices measure the momentum and the energy loss, $dE/dx$,
of the tracks.  The DIRC provides a good pion-kaon separation by
measuring the angle of the Cherenkov light emitted by the particles. The
amount of deposited energy and the shape of showers induced in the EMC
are used to distinguish between electrons, muons and hadrons. The energy
deposits in the IFR are used to distinguish between muons and hadrons.

To reduce discrepancies between simulated and real data, we require that
the Cherenkov angle in the DIRC be consistent with the momentum of the
kaon candidates in the laboratory frame.  Each track is tested
sequentially for identification as muon, electron, kaon and pion, and is
classified according to the first successfull identification, or as a
non-identified track if all identifications fail.  The efficiencies of
PID requirements are measured on data samples by \babar.

A signal candidate event must have one track identified as an electron
or a muon, and the other one identified as either a kaon or a pion. The
presence of an identified lepton and hadron defines the tag and signal
hemisphere, respectively. The hadron track is required not to exceed
3.5\gevc, in order to suppress di-lepton background, while the momentum
of all leptons and the momentum of the pion in the \taumtopi mode has to
be larger than 1\gevc in the laboratory frame, to reduce particle
misidentification rates. Events with additional tracks are discarded.

Photon candidates are reconstructed using well-formed EMC clusters with
an energy $E>75\mev$ and not associated with a track.  Photon pairs are
combined to form \piz candidates if they have an invariant mass
$90 < m_{\gamma \gamma} <165 \mevcc$. If two candidates share an EMC
cluster, only the candidate with $m_{\gamma\gamma}$ closer to the \piz
mass $m_{\piz}= 134.977\gevcc$~\cite{Patrignani:2016xqp} is selected to
avoid double counting. The \piz candidates are required to have an
energy in the laboratory frame of at least 200\mev, and to fly with a
angle smaller than $1.5\rad$ with respect to the signal charged
particle.  To reduce background and cross-feed contamination, we discard
events containing any additional photon that has momentum direction
within $1.5\rad$ with respect to the signal track and cannot be paired
to reconstruct a \piz candidate. This requirement is referred to as
``extra photon veto''.

The thrust magnitude must be smaller than 0.99 and the angle between the
two track must be smaller than $2.95\rad$.  The missing mass of the
event is computed subtracting the event candidate 4-momentum from the
CM-energy 4-momentum and is required to be larger than 1.0\gevcc for
$n_{\piz} >0$, and larger than 2.5\gevcc for $n_{\piz} =0$, where
$n_{\piz}$ is the number of reconstructed \piz's. These last three
requirements suppres radiative Bhabha and di-muon backgrounds.

Two-photon events, in which the final-state $e^-$ and $e^+$ are
scattered at small angles outside the detector acceptance, are removed
by requiring a missing mass smaller than 7.5\gevcc. For events with
$n_{\piz} =0$, we also require that the ratio of the transverse momentum
in the event, $p_T$, and the missing energy,
$E_{\text{miss}} = \sqrt{s} - p_{\text{tag}} - p_{\text{sig}}$, be
$>0.2$, where $p_{\text{tag}}$ and $p_{\text{sig}}$ are the moduli of the
momenta of the tag and signal tracks, respectively.

We suppress backgrounds from events with undetected $K_L$'s or with
spurious extra reconstructed particles by requiring that the signal
hemisphere missing mass is within decay-mode-dependent limits. To
compute the missing mass, the signal \mtau energy is set to one half the
CM energy and its momentum direction is set to the thrust direction.

According to simulation, the selection efficiency ranges from 0.13\%
(for \taumtoKthreepiz{}) to 3.3\% (for \taumtoKpiz{}), and the fraction
of background and cross-feed ranges from 5.5\% (for \taumtopipiz{}) to
79\% (for \taumtoKthreepiz{}).

\section{Systematics studies}
\label{sec:syst-studies}

For the simulation of the PID efficiencies, we use the \babar\ PID
efficiencies measurements in all cases except for the efficiencies to
identify a pion as a pion, a kaon as a kaon and a pion as a kaon. We
determine these three efficiencies using 3-prong \mtau decay modes
\taumtopipipi and \taumtoKKpi, following a strategy similar to
Ref.~\cite{Aubert:2009qj}. These control samples have a low
charged-particle multiplicity similar to the signal modes and are
selected in events with a 1-3 prong topology, where the charged particle
in the 1-prong hemisphere is identified as an electron or muon. The
selection requirements are as close as possible to the ones used for the
selection of the signal and control modes.  We obtain an unbiased high
purity \Km sample by selecting candidate decays \taumtoKKpi where we
identify the \Kp and the \pim.  The remaining particle has to be a \Km
with high probability rather than a \pim, in order to be consistent with
the hadronization of the virtual $W^-$ that mediates the \taum decay.
Similarly, we select an unbiased high purity sample of \pip's in
\taumtopipipi decays where we identify both \pim's.  We use the pure \Km
and \pip samples to measure the above mentioned three PID efficiencies
as a function of the \babar\ data taking period, the particle charge and
true type, momentum, and polar and azimuthal angles.

\begin{figure}
  \centering \sidecaption \raisebox{\height}{%
    \begin{tabular}{@{}c@{}}
      \begin{overpic}[width=0.5\linewidth, trim=0 0 0 0, clip]{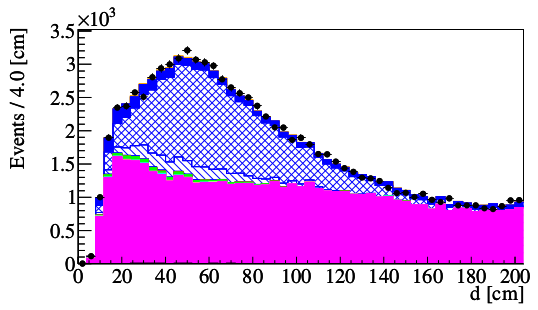}
        \put(80,40){(a)}
      \end{overpic}
      \\
      \begin{overpic}[width=0.5\linewidth, trim=0 0 0 0, clip]{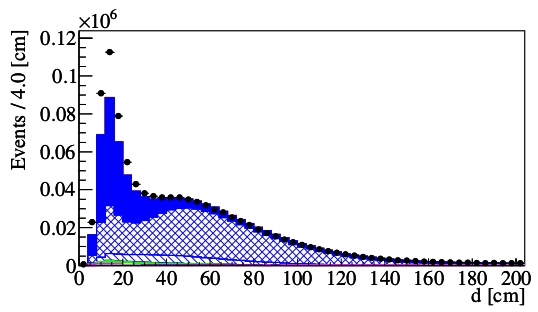}
        \put(80,40){(b)}
      \end{overpic}
    \end{tabular}%
  }
  \caption{Distance $d$ between the track intersection point with the
    EMC and the cluster centroid of the closest reconstructed
    photon. Plot (a) reports \taumtomu candidates, plot (b)
    reports \taumtopi candidates. Data points are overlaid onto
    cumulated histograms representing simulated samples, drawn with the
    patterns documented in Figure~\ref{fig:lab-momentum-signal}.  }
  \label{fig:splitoffDist}
\end{figure}

Charged hadron showers in the EMC may include neutrons than further
interact with the EMC at some distance, producing separate (split-off)
showers that are not associated with a track and can be reconstructed as
photon candidates. The reliability of the Monte Carlo simulation of
these fake split-off photons has been studied with data and simulated
control samples of candidate \taumtomu and \taumtopi decays.  These
samples have been selected in the same way as the signal samples, except
that for the \taumtomu sample the other track is required to be an
identified electron rather that either an eletron or a muon.  While the
simulation accurately describes the reconstructed photons in the signal
hemisphere for muon tracks, the data events with pion tracks exhibit a
significant excess of photon candidates corresponding to EMC energy
deposits located within 40\,cm of the track-EMC intersection, as
illustrated in Figure~\ref{fig:splitoffDist}. The measured excess of
reconstructed photons is used to compute a correction weight of
$\eta_{\text{so}} = 0.972$ for the simulated efficiency of the extra
photon veto requirement for the signal events with either a pion of a
kaon.

\begin{figure}
  \centering
  \sidecaption
  \includegraphics[width=0.5\linewidth,clip]{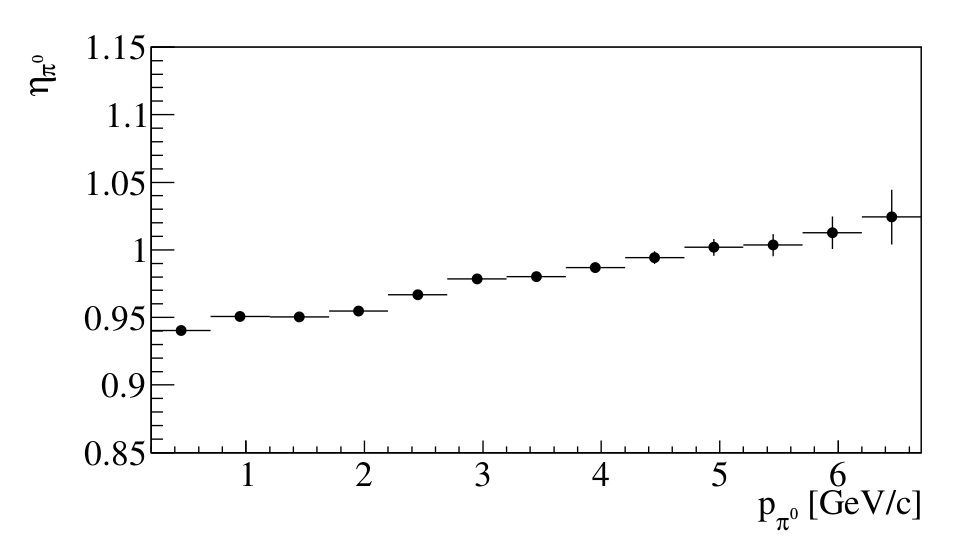}
  \caption{Correction weights for the \piz reconstruction efficiency
    as a function of the \piz momentum $p_{\piz}$.}
  \label{fig:pi0_eff_corr}
\end{figure}

The accuracy of the Monte Carlo simulation of the \piz reconstruction
efficiency has been studied on data and simulated control samples
containing \mtau decays to one track and zero, one or two \piz's
[$\taumtotnpiz$ $(n=0,1,2)$], which have been selected as the signal
samples, accepting any signal track that is not an identified electron,
and requiring an identified electron in the tag hemisphere.  As a
result, the signal track $t^-$ can be either a muon, a pion or a
kaon candidate. An \piz-momentum-dependent correction weight for the simulated
\piz reconstruction efficiency is obtained by comparing the data and
simulated ratio of events with one and zero reconstructed
\piz's. Its value is shown in Figure~\ref{fig:pi0_eff_corr}.
Averaged on the \piz momentum, the correction weight is
$\eta_{\piz} = 0.958 \pm 0.001\,\stat \pm 0.009\,\syst$, where the
statistical uncertainty is given by the sample sizes and the systematic
uncertainty is determined by the uncertainty on the split-off
correction, the uncertainties on the \mtau branching fractions used in
the simulation and the uncertainty on the estimate of the amount of
Bhabha background in the control samples. When using the above
correction weights, the simulated momentum distribution of the
reconstructed \piz's matches the data within statistical uncertainties
both on the sample with one reconstructed \piz that has been used to
obtain the weights and on the independent sample with two reconstructed
\piz's.

\begin{figure}
  \centering
  \begin{tabular}{@{}c@{}c@{}}

    \begin{overpic}[trim=0 0 30 -25, width=0.5\linewidth]{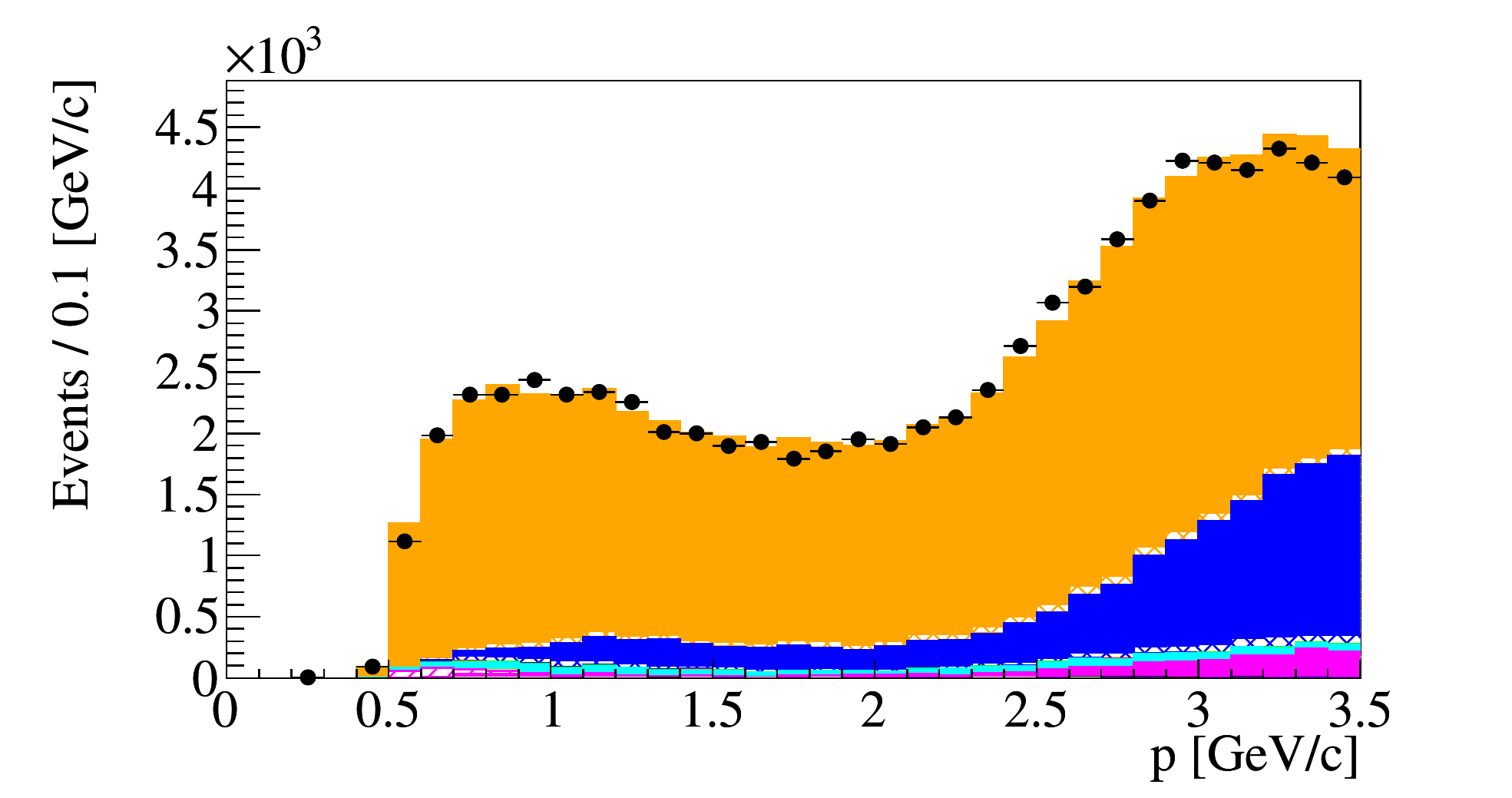}
      \put(30,51){\smaller\smaller\bfseries$\tau^- \to K^- \nu_\tau$}
      \put(45,39){
        \begin{varwidth}{\linewidth}\smaller\smaller\smaller
          \babar\\[-0.5ex]
          preliminary
        \end{varwidth}}
    \end{overpic}
    & 
    \begin{overpic}[trim=0 0 30 -25, width=0.5\linewidth]{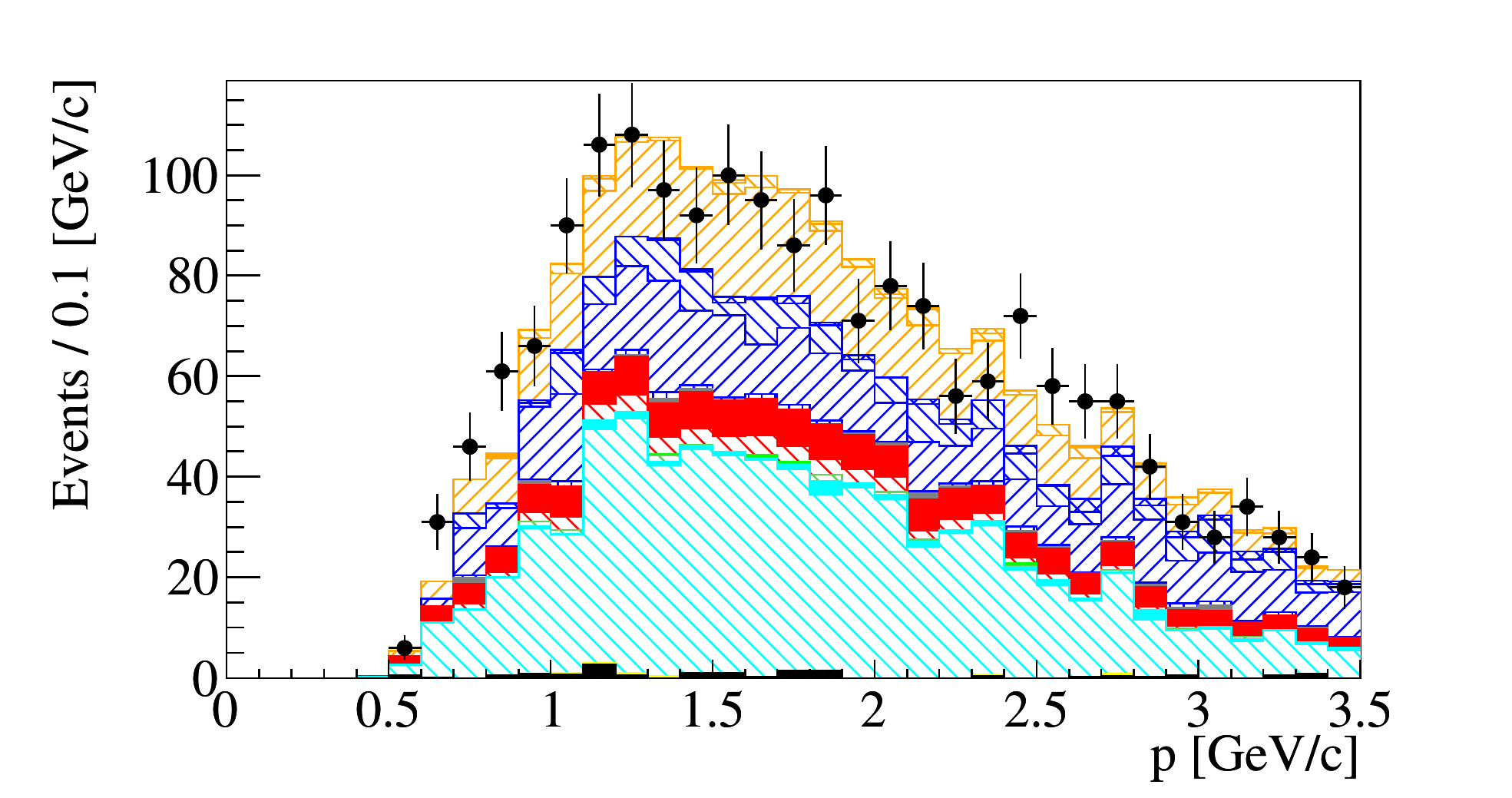}
      \put(30,51){\smaller\smaller\bfseries$\tau^- \to K^- 3\pi^0 \nu_\tau$}
      \put(65,39){
        \begin{varwidth}{\linewidth}\smaller\smaller\smaller
          \babar\\[-0.5ex]
          preliminary
        \end{varwidth}}
    \end{overpic}
    \\[-2ex]
    \begin{overpic}[trim=0 0 30 -25, width=0.5\linewidth]{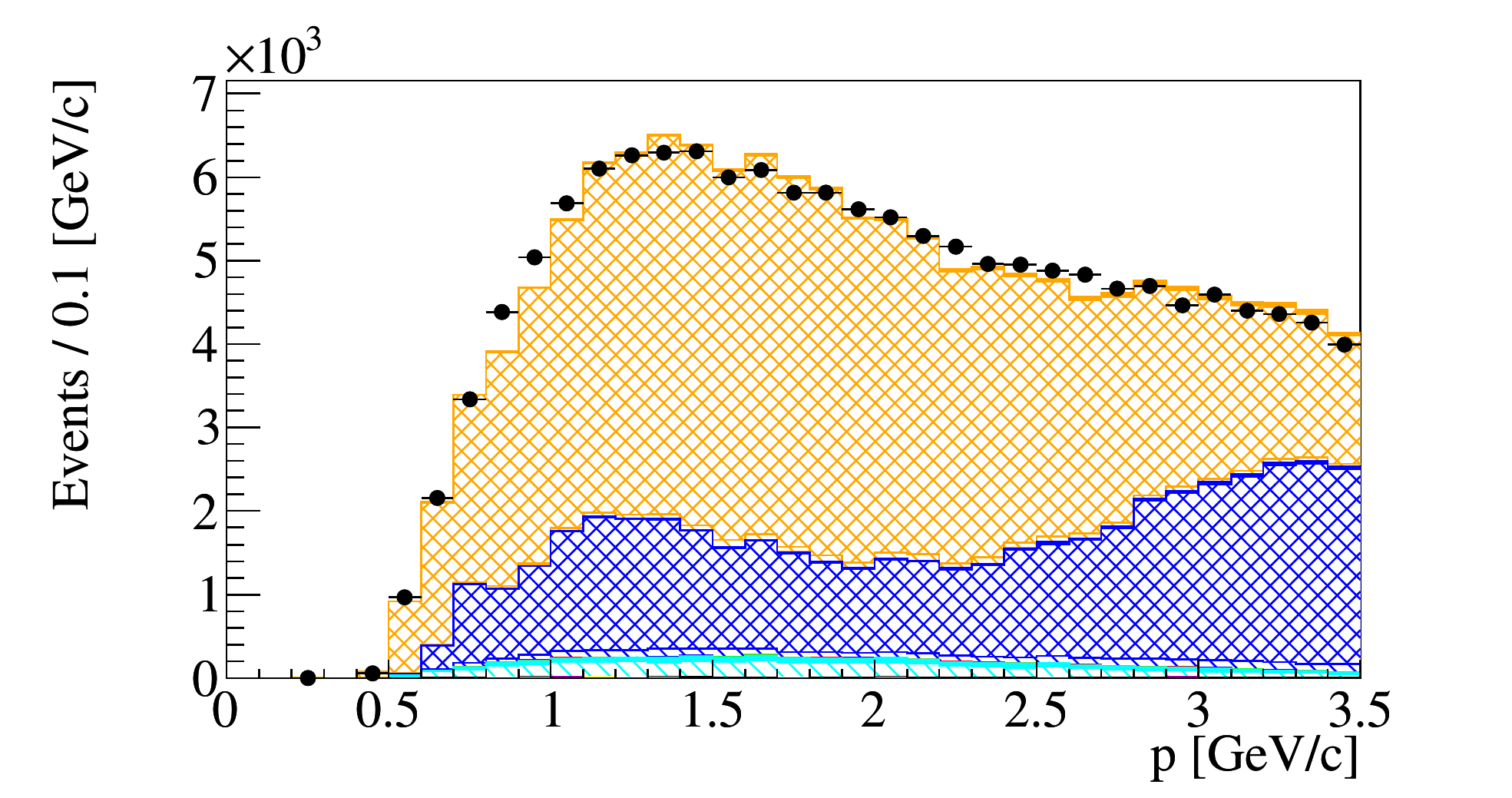}
      \put(30,51){\smaller\smaller\bfseries$\tau^- \to K^- \pi^0 \nu_\tau$}
      \put(65,39){
        \begin{varwidth}{\linewidth}\smaller\smaller\smaller
          \babar\\[-0.5ex]
          preliminary
        \end{varwidth}}
    \end{overpic}
    &
    \begin{overpic}[trim=0 0 30 -25, clip, width=0.5\linewidth]{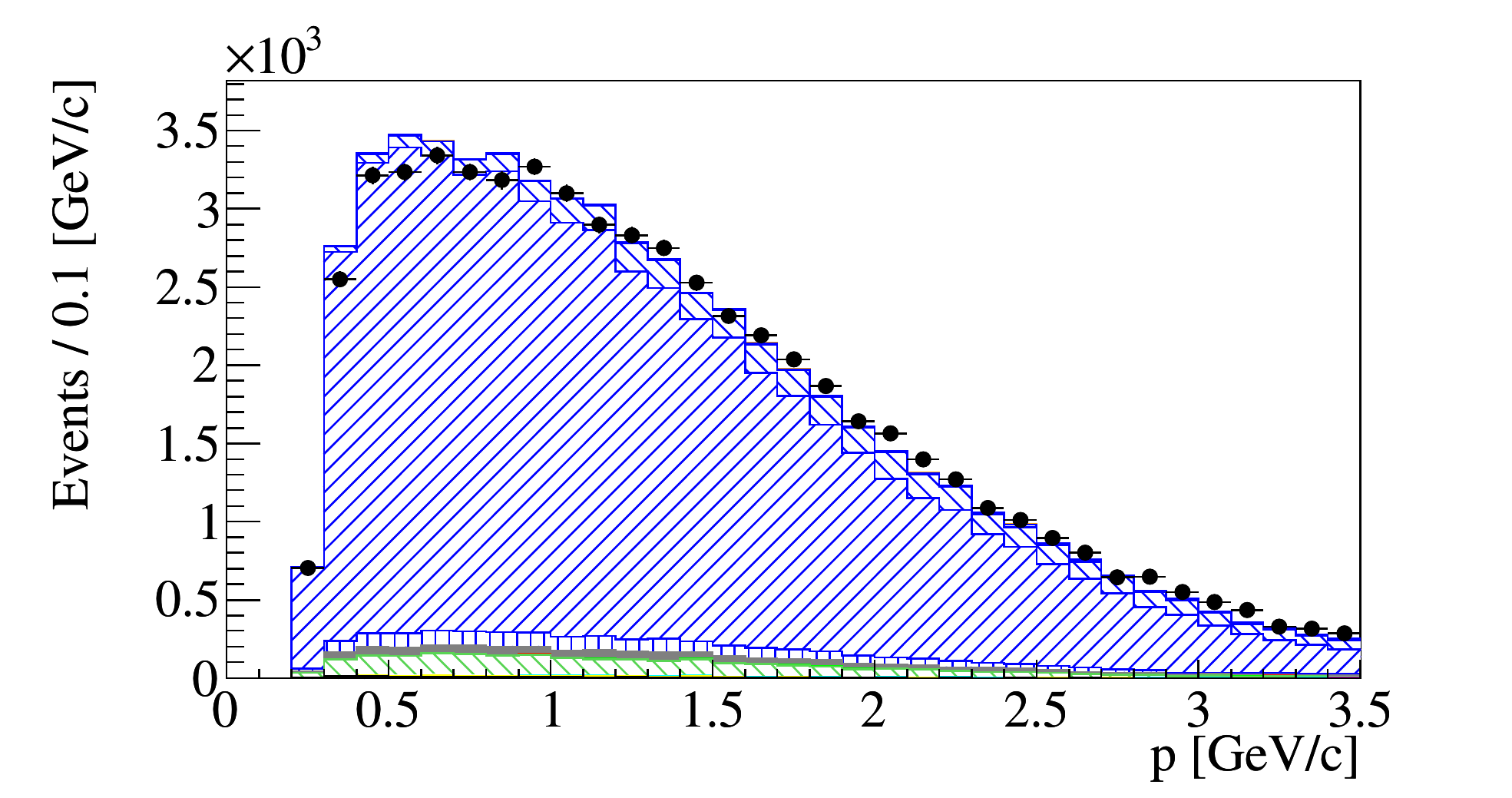}
      \put(30,51){\smaller\smaller\bfseries$\tau^- \to \pi^- 3\pi^0 \nu_\tau$}
      \put(65,39){
        \begin{varwidth}{\linewidth}\smaller\smaller\smaller
          \babar\\[-0.5ex]
          preliminary
        \end{varwidth}}
    \end{overpic}
    \\[-2ex]
    \begin{overpic}[trim=0 0 30 -25, clip, width=0.5\linewidth]{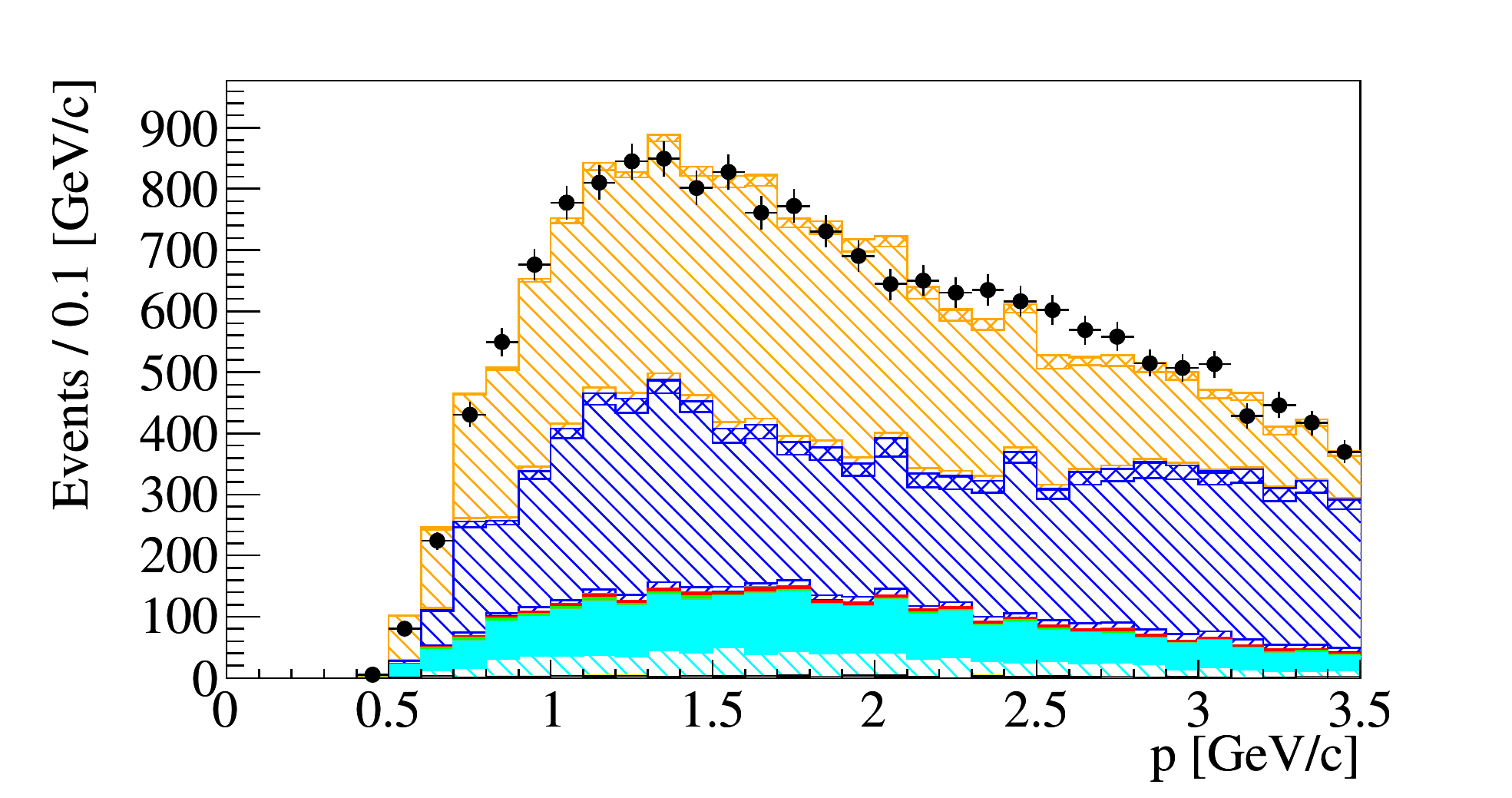}
      \put(30,51){\smaller\smaller\bfseries$\tau^- \to K^- 2\pi^0 \nu_\tau$}
      \put(65,39){
        \begin{varwidth}{\linewidth}\smaller\smaller\smaller
          \babar\\[-0.5ex]
          preliminary
        \end{varwidth}}
    \end{overpic}
    &
    \begin{overpic}[trim=0 0 30 -25, clip, width=0.5\linewidth]{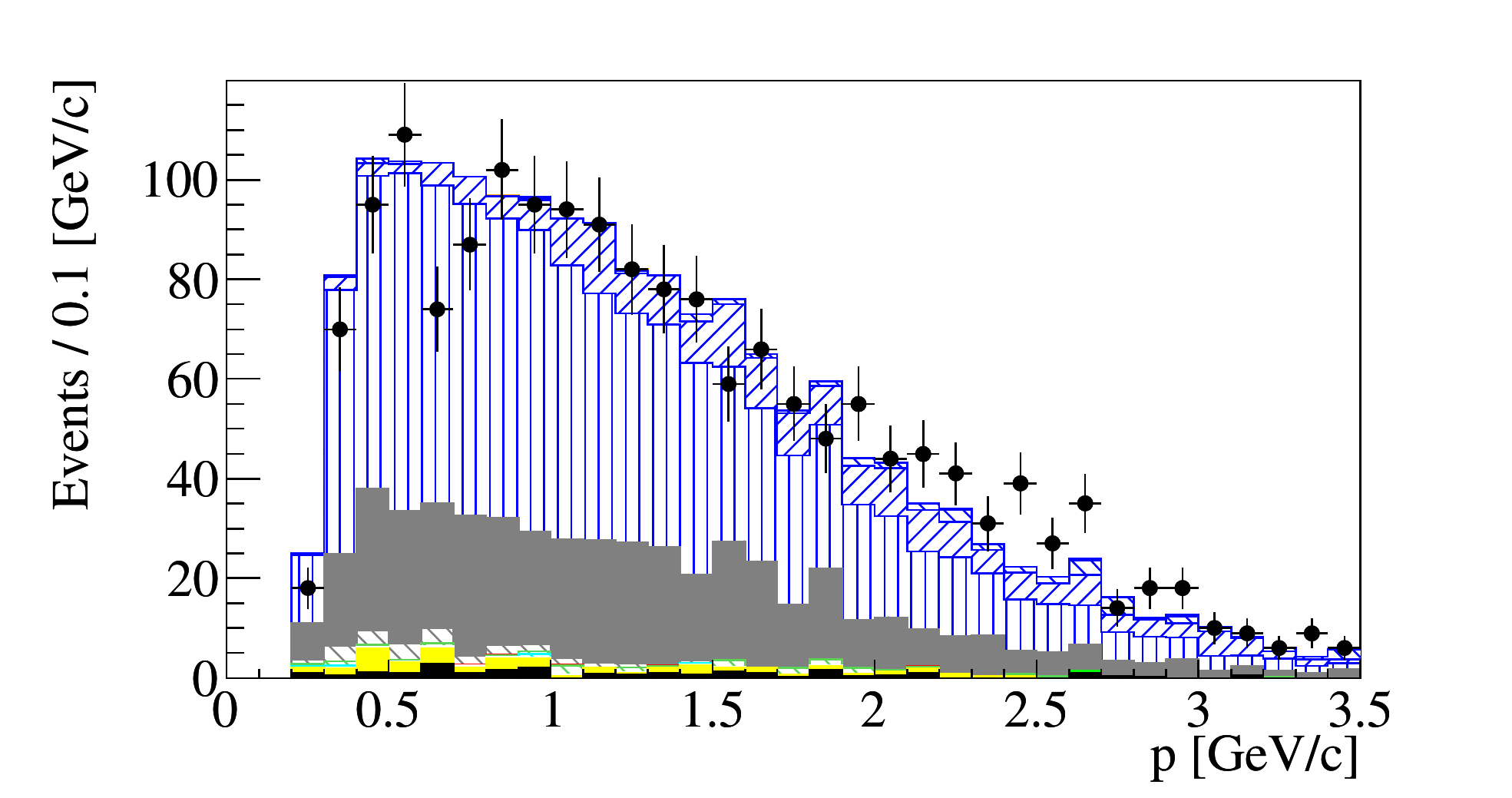}
      \put(30,51){\smaller\smaller\bfseries$\tau^- \to \pi^- 4\pi^0 \nu_\tau$}
      \put(65,39){
        \begin{varwidth}{\linewidth}\smaller\smaller\smaller
          \babar\\[-0.5ex]
          preliminary
        \end{varwidth}}
    \end{overpic}
    \\
    \multicolumn{2}{c}{%
      \begin{overpic}[width=0.9\linewidth]{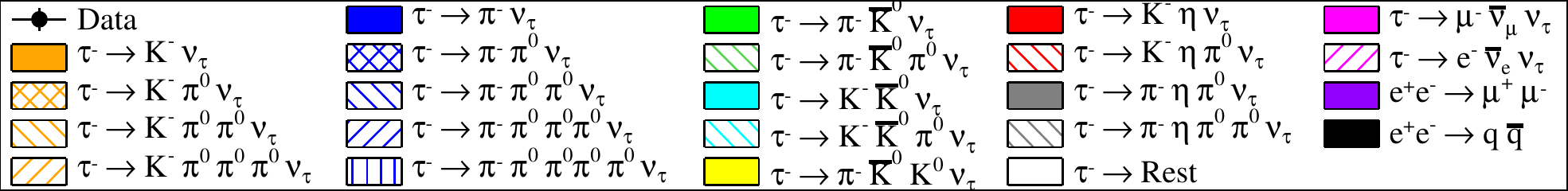}
      \end{overpic}%
    }
  \end{tabular}
  \caption{Laboratory-frame momentum of the track in the signal hemisphere for the
    selected candidates of the six signal modes. Data points are overlaid onto cumulated histograms
    representing simulated samples.}
  \label{fig:lab-momentum-signal}
\end{figure}
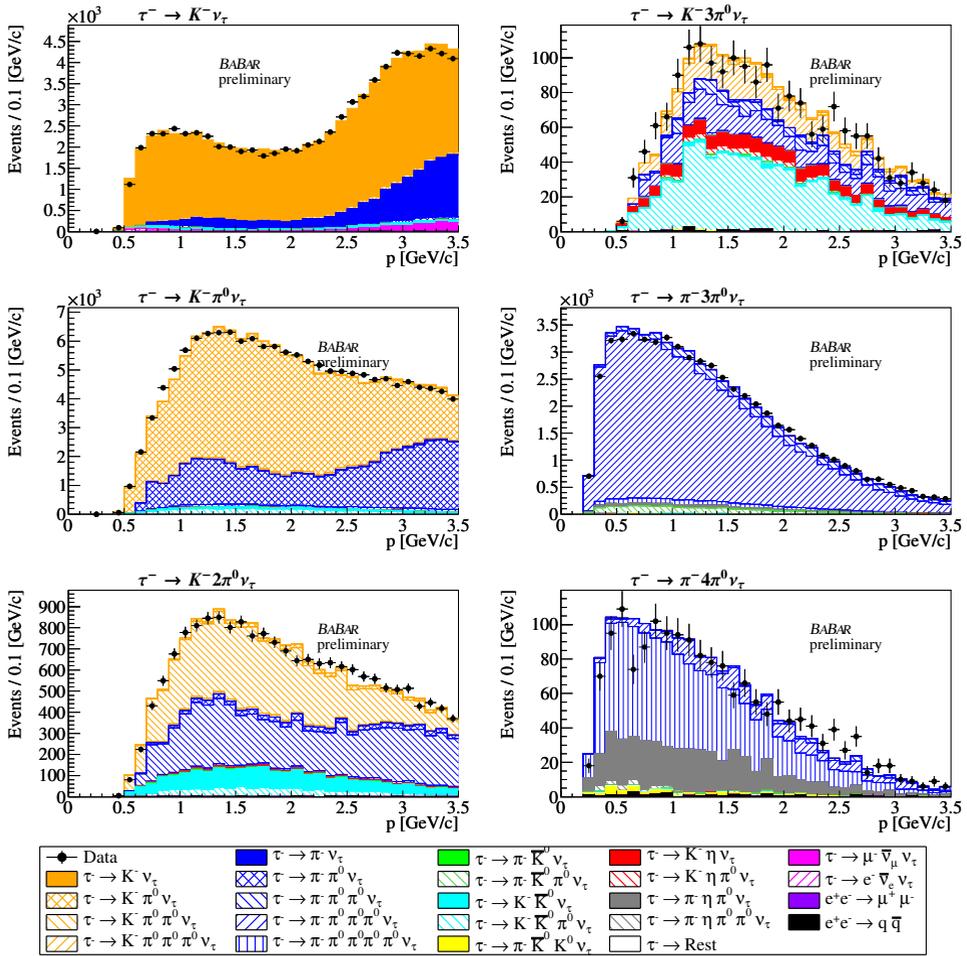

Figure~\ref{fig:lab-momentum-signal} shows that, after applying all
corrections, and after using in the simulation also the branching
fractions that are measured in this analysis, the simulation of the
signal track momentum in the laboratory frame reproduces the data quite
accurately for all the signal modes.

\section{Determination of the branching fractions}
\label{sec:br-meas}

The selected candidates include backgrounds from the other signal modes
(cross-feed) and from events other than the signal modes. These latter
backgrounds are subtracted using the Monte Carlo simulation of
electron-positron annihilations to pairs of muons, \mtau leptons and to
final states of light quarks ($uds$), charm and bottom hadrons.
Background contributions from Bhabha and two-photon events are estimated to
be negligible on data.  Cross-feed backgrounds are subtracted by inverting the
matrix $M_{ij}$ that describes the selection efficiency of
reconstructing an event containing one or two decays of the signal mode
$i$ into any signal candidate sample $j$. $M_{ij}$ is measured on
simulated events. Thus:
\begin{align}
  N_{i}^{\text{Prod}} &= \sum_{j} (M^{-1})_{ij}\left( N_{j}^{\text{Sel}} - N_{j}^{\text{Bkg}}\right)~,
  \label{eq:crossfeed-removal}
\end{align}
where, for each signal mode $i$, $N_i^{\text{Prod}}$ denotes the
efficiency-corrected number of produced events, while $N_i^{\text{Sel}}$
and $N_i^{\text{Bkg}}$ denote the numbers of selected candidates and of
estimated background events, respectively.  The branching fractions are
then:
\begin{align}
  \BR(\mtau \to i) =& 1 - \sqrt{ 1 - 2\frac{N_i^{\text{Prod}}}{ N_{\tau} }}~,
  \label{eq:br}
\end{align}
where $N_{\tau} = 2\L \sigma_{\tau\tau}$ is the the number of produced
\mtau leptons, obtained from the estimate of the integrated luminosity
corresponding to the analyzed data sample, \L~\cite{Lees:2013rw}, and
the $e^+e^-\to\tau^+\tau^-$ cross-section
$\sigma_{\tau\tau}$~\cite{Banerjee:2007is} at and around the
$\Upsilon(4S)$ peak. The expression in Eq.~\ref{eq:br} originates from
the choice to include in $N_i^{\text{Prod}}$ events with both one or two
signal-mode-$i$ \mtau decays. The statistical uncertainties on the number of the
signal samples' candidates are determined by the samples' sizes and are
independent from each other. Eq.~\ref{eq:crossfeed-removal} and
\ref{eq:br} determine how the statistical covariance matrix of the branching
fractions is computed from the signal-candidates samples' uncertainties.
The signal branching fractions' values and statistical uncertainties are reported
on Table~\ref{tab:systErrorSignal}, and their statistical correlation is
reported on Table~\ref{tab:statCorr}.

\section{Systematic uncertainties}
\label{sec:syst-unc}

The contribution to the systematic covariance matrix of the signal
branching fractions from the uncertainty on a quantity $X_i$ are
computed by varying 50 times $X_i$ according to a Gaussian distribution
and by recomputing all signal branching fractions for each variation.
The contributions to the total systematic systematic uncenrtainties on
the signal branching fractions are reported in
Table~\ref{tab:systErrorSignal}, while the total systematic correlation
is reported on Table~\ref{tab:systCorr}.

The coefficients of the efficiency and mixing matrix $M_{ij}$ in
Eq.~\ref{eq:crossfeed-removal} have uncertainties determined by the
uncertainties on simulated selection efficiencies. We express the
uncertainties on the $M_{ij}$ coefficients as a function of independent
statistical uncertainties of the selected samples in the simulation, and
we compute an overall $M_{ij}$ contribution to the systematic covariance
of the branching fractions by summing all contributions from these
independent uncertainties. In the following, this systematic
contribution is referred to as ``Signal efficiencies'' contribution.

The systematic contribution due to the finite size of the simulation
samples used to estimate the selection efficiencies of the background
contaminations are calculated using the number of events in the involved samples.

For background subtraction, the simulation relies on the PDG
2017~\cite{Patrignani:2016xqp} averages of the \mtau branching
fractions.  We
vary those branching fractions independently according to their
uncertainties to estimate the induced systematic contributions on the
measurements.  \iffalse +++ The largest systematic uncertainty
contribution is found for the \taumtoKthreepiz mode,
$\Delta \BR = 35\pc$. The main source of this uncertainty is the large
background contribution ($30\pc$) from the \taumtoKKzpiz decays, whose
relative uncertainty is $16\pc$.  \else The largest systematic
uncertainty contribution is found for the \taumtoKthreepiz mode and is
due to the subtraction of a large background contamination from
\taumtoKKzpiz decays, whose branching fraction is not well known.\fi

The decays \taumtopifivepiz and \taumtoKfourpiz are not included in the
background simulation. We estimate a systematic contribution due to the
omission of these modes in the simuation and hence in the background
subtraction by selecting candidates for these modes in data and in the
simulation. All selected candidates in the simulation are necessarily
background. We estimate the selection efficiency using the respective
samples with one-less \piz and the measured \piz efficiency for the
additional \piz. We compute 68\% CL upper limits on the presence of
these decay modes in data, and we use the measured \piz reconstruction
inefficiency to estimate the corresponding backgroung contributions to
the selected signal-candidates samples. We compute the systematic
uncertainties by varying the background contaminations around zero with
an uncertainty equal to the respective 68\% CL upper limits.

The estimated number of produced \mtau decays in data, $N_\tau$, is used
in Eq.~\ref{eq:br} and to weight the events of simulated samples for
background subtraction to match the data. $N_\tau$ is varied according
to the uncertainties on the integrated luminosity of the data sample and
on $\sigma(e^+e^- \to \tau^+\tau^-)$ to compute the associated
systematics.

\begin{table*}[tb]
  \centering
  \caption{Summary of the preliminary measured branching fractions and their
    uncertainties. Uncertainties that are relative to their branching
    fraction value are reported as percentages and labelled with
    ``[\%]''. The total uncertainty is obtained by adding the
    statistical and systematic uncertainties in
    quadrature.}
  \label{tab:systErrorSignal}
  \begin{tabular}{lrrrrrr}
    \toprule
    Decay mode & \multicolumn{1}{c}{\dK} & \multicolumn{1}{c}{\dKpiz} & \multicolumn{1}{c}{\dKtwopiz} &
    \multicolumn{1}{c}{\dKthreepiz} & \multicolumn{1}{c}{\dpithreepiz} & \multicolumn{1}{c}{\dpifourpiz}
    \\
    &  \smaller$(\times 10^{-3})$ & \smaller$(\times 10^{-3})$ & \smaller$(\times 10^{-4})$ &
    \smaller $(\times 10^{-4})$ & \smaller $(\times 10^{-2})$ & \smaller $(\times 10^{-4})$
    \\
    \midrule
    Branching fraction & \htuse{central_val_abs} \\
    Stat. uncertainty & \htuse{stat_uncert_abs} \\
    Syst. uncertainty & \htuse{syst_uncert_abs} \\
    Total uncertainty & \htuse{tot_uncert_abs} \\
    
    \hline\rule{0pt}{2.5ex}%
    
    Stat. uncertainty [$\%$]& \htuse{stat_uncert} \\
    Syst. uncertainty [$\%$]& \htuse{syst_uncert} \\
    Total uncertainty [$\%$]& \htuse{tot_uncert} \\
    
    \hline\rule{0pt}{2.5ex}%

    Signal efficiencies [$\%$]        & \htuse{signalEffi} \\

    Background efficiency [$\%$]      & \htuse{bkgEffi} \\

    MC \mtau branching fractions [$\%$]            & \htuse{bkgNorm} \\
    $\pi5\piz$ background [$\%$]& \htuse{downFeedP} \\
    $K4\piz$ background [$\%$]  & \htuse{downFeedK} \\

    Number of \mtau decays [$\%$]   & \htuse{lumi} \\

    \babar\ PID [$\%$]                & \htuse{pidTable} \\
    Custom PID [$\%$]                 & \htuse{pidCorr} \\
    Muon mis-id [$\%$]                & \htuse{muMisID} \\
    Track efficiency [$\%$]          & \htuse{trackEffi} \\
    Split-off correction [$\%$]      & \htuse{splitoff} \\
    \piz correction  [$\%$]          & \htuse{pi0Cor}    \\
    \bottomrule
  \end{tabular}
\end{table*}

\begin{table}
  \centering
  \caption{Statistical correlation matrix for the branching fractions of
    the signal modes (preliminary).}
  \label{tab:statCorr}
  \begin{tabular}{lrrrrrr}
    \toprule
    \htuse{stat_corr_mat}
    \bottomrule
  \end{tabular}
\end{table}
\begin{table}
  \centering
  \caption{Systematic correlation matrix for the branching fractions of
    the signal modes (preliminary).}
  \label{tab:systCorr}
  \begin{tabular}{lrrrrrr}
    \toprule
    \htuse{syst_corr_mat}
    \bottomrule
  \end{tabular}
\end{table}
\begin{table}
  \centering
  \caption{Total correlation matrix for the branching fractions of the
    signal modes (preliminary).}
  \label{tab:totCorr}
  \begin{tabular}{lrrrrrr}
    \toprule
    \htuse{tot_corr_mat}
    \bottomrule
  \end{tabular}
\end{table}

The \babar PID selectors efficiencies are varied according to their
uncertainties to obtain their systematic contribution, labelled ``\babar
PID''. The PID efficiencies measured with the dedicated study performed
for this analysis are also varied to get the contribution labelled
``custom PID''. To account for discrepancies between the data and the
simulation, the efficiency of identifying a true muon as a pion or a
kaon is varied by 50\%. The associated systematic contribution is
non-negligible only for the \taumtoK decay mode.

Systematic uncertainties in simulating the tracking efficiencies have
been estimated by \babar using data control
samples~\cite{Allmendinger:2012ch} and amount to 0.17\%.  These
uncertainties are assumed to be fully correlated for the 2 tracks in all
signal modes. The selected data events are weighted with random weights
centered on 1 and with 0.17\% uncertainty to compute the associated
systematics.

To get the corresponding systematics, we vary the correction weight of
$\eta_{\text{so}} = 0.972$ that is applied on simulated events to adjust
for the insufficient production of split-off photons on simulated events
with hadronic tracks, using an uncertainty of 50\% of its deviation from
1. The uncertainty on the correction weight due to the sample sizes is
comparatively negligible.

The \piz-momentum-dependent weights that adjust the simulation to the
data regarding the \piz reconstruction efficiencies are all coherently
varied according to the total uncertainty on the momentum-averaged
correction weight,
$\eta_{\piz} = 0.958 \pm 0.001\,\stat \pm 0.009\,\syst$.

\section{Results}
\label{sec:results}

Using the data sample of $435.5 \times 10^6$ \mtau-pairs recorded by the
\babar experiment, we provide preliminary measurements of the following
six \mtau decay branching fractions, excluding contributions proceeding
through $K^0$ and $\eta$ mesons:
\begin{align*}
  & \BR(\taumtoK)         & ={} & \htuse{BrK_statsyst},\\
  & \BR(\taumtoKpiz)      & ={} & \htuse{BrKPi0_statsyst}, \\
  & \BR(\taumtoKtwopiz)   & ={} &  \htuse{BrK2Pi0_statsyst}, \\
  & \BR(\taumtoKthreepiz) & ={} & \htuse{BrK3Pi0_statsyst}, \\
  & \BR(\taumtopithreepiz)& ={}  & \htuse{BrPi3Pi0_statsyst}, \\
  & \BR(\taumtopifourpiz) & ={} & \htuse{BrPi4Pi0_statsyst},
\end{align*}
where the first uncertainty is statistical and the second one is
systematic. The correlation matrices of the statistical, systematic and
total uncertainties are reported in Tables~\ref{tab:statCorr},
\ref{tab:systCorr}, and \ref{tab:totCorr}, respectively.

The result for $\BR(\taumtoK)$ is consistent with an earlier \babar
measurement~\cite{Aubert:2009qj}, which used a different tagging
technique (3-prong hadronic tag) and thus relies on a statistically
independent data sample.  The result for \taumtoKpiz is meant to
eventually supersede an earlier \babar measurement~\cite{Aubert:2007jh},
which shares part of the sample of this analysis, has a less
sophisticated treatment of systematic effects, and deviates by
$3.8\sigma$ from this paper measurement, when assuming that the old and
new uncertainties are fully uncorrelated.

The measured branching fractions with kaons have significantly
improved precision compared to earlier measurements at LEP and at
Cornell, and are consistent with those
results~\cite{Patrignani:2016xqp}.

\ifdefined\bibtexflag
\bibliography{%
  lusiani-phipsi19-procs.bib%
  ,bibtex/pub-2018%
  ,bibtex/pub-2017%
  ,bibtex/pub-2016%
  ,bibtex/pub-2014%
  ,bibtex/pub-2013%
  ,bibtex/tau-lepton%
  ,bibtex/pub-extra%
  ,bibtex/pub-extra-noauthor%
  ,bibtex/misc%
  ,bibtex/biblio-superb%
  ,tau-refs%
  ,tau-refs-pdg%
  ,lusiani-ichep16-babar-bib%
} \else

\begin{thebibliography}{20}

\bibitem{Gamiz:2002nu}
E.~Gamiz, M.~Jamin, A.~Pich, J.~Prades, F.~Schwab, JHEP \textbf{01}, 060
  (2003), \texttt{hep-ph/0212230}

\bibitem{Gamiz:2004ar}
E.~Gamiz, M.~Jamin, A.~Pich, J.~Prades, F.~Schwab, Phys. Rev. Lett.
  \textbf{94}, 011803 (2005), \texttt{hep-ph/0408044}

\bibitem{Amhis:2016xyh}
Y.~Amhis et~al. (HFLAV), Eur. Phys. J. \textbf{C77}, 895 (2017),
  \texttt{1612.07233}

\bibitem{Patrignani:2016xqp}
C.~Patrignani et~al. (Particle Data Group), Chin. Phys. \textbf{C40}, 100001
  (2016)

\bibitem{Lusiani:2018zvr}
A.~Lusiani, \emph{{Status and progress of the HFLAV-Tau group activities}}
  (2018), {to appear in the proceedings of the 'International Workshop on
  e$^+$e$^-$ collisions from Phi to Psi, Mainz, Germany}, \texttt{1804.08436},
  \urlstyle{tt}\url{https://inspirehep.net/record/1669594/files/1804.08436.pdf}

\bibitem{Aubert:2001tu}
B.~Aubert et~al. (BaBar), Nucl. Instrum. Meth. \textbf{A479}, 1 (2002),
  \texttt{hep-ex/0105044}

\bibitem{Jadach:1999vf}
S.~Jadach, B.F.L. Ward, Z.~Was, Comput. Phys. Commun. \textbf{130}, 260 (2000),
  \texttt{hep-ph/9912214}

\bibitem{Banerjee:2007is}
S.~Banerjee, B.~Pietrzyk, J.M. Roney, Z.~Was, Phys. Rev. \textbf{D77}, 054012
  (2008), \texttt{0706.3235}

\bibitem{TheBABAR:2013jta}
B.~Aubert et~al. ({BaBar}), Nucl. Instrum. Meth. \textbf{A729}, 615 (2013),
  \texttt{1305.3560}

\bibitem{Jadach:1993hs}
S.~Jadach, Z.~Was, R.~Decker, J.H. Kuhn, Comput. Phys. Commun. \textbf{76}, 361
  (1993)

\bibitem{Sjostrand:1993yb}
T.~Sjostrand, Comput.Phys.Commun. \textbf{82}, 74 (1994)

\bibitem{Lange:2001uf}
D.~Lange, Nucl.Instrum.Meth. \textbf{A462}, 152 (2001)

\bibitem{Golonka:2005pn}
P.~Golonka, Z.~Was, Eur. Phys. J. \textbf{C45}, 97 (2006),
  \texttt{hep-ph/0506026}

\bibitem{Agostinelli:2002hh}
S.~Agostinelli et~al. (GEANT4), Nucl.Instrum.Meth. \textbf{A506}, 250 (2003)

\bibitem{Allison:2006ve}
J.~Allison, K.~Amako, J.~Apostolakis, H.~Araujo, P.~Dubois et~al. (Geant4
  collaboration), IEEE Trans.Nucl.Sci. \textbf{53}, 270 (2006)

\bibitem{Brandt:1964sa}
S.~Brandt, C.~Peyrou, R.~Sosnowski, A.~Wroblewski, Phys. Lett. \textbf{12}, 57
  (1964)

\bibitem{Aubert:2009qj}
B.~Aubert et~al. (\babar), Phys. Rev. Lett. \textbf{105}, 051602 (2010),
  \texttt{0912.0242}

\bibitem{Lees:2013rw}
J.P. Lees et~al. ({BaBar}), Nucl. Instrum. Meth. \textbf{A726}, 203 (2013),
  \texttt{1301.2703}

\bibitem{Allmendinger:2012ch}
T.~Allmendinger et~al., Nucl. Instrum. Meth. \textbf{A704}, 44 (2013),
  \texttt{1207.2849}

\bibitem{Aubert:2007jh}
B.~Aubert et~al. (\babar), Phys. Rev. \textbf{D76}, 051104 (2007),
  \texttt{0707.2922}

\end{thebibliography}

 \fi
\end{document}